\documentclass[epj]{svjour}

\usepackage{graphics}

\newcommand{\Name}[1]{#1, }
\newcommand{\REVIEW}[4]{#1 {\bf #2}, (#3) #4.}
\newcommand{\etal}{{\it et al.\ }}
\begin{document}

\title{
Tunable-slip boundaries for coarse-grained simulations of fluid flow
}

\author{J. Smiatek\inst{1} \and M.\ P. Allen\inst{2} \and F. Schmid\inst{1}}

\institute{
  \inst{1} Physics Faculty, Universit\"at Bielefeld, D-33615 Bielefeld, Germany\\
  \inst{2} Department of Physics and Centre for Scientific Computing,
   University of Warwick, Coventry CV4 7A1, UK
}

\abstract{ On the micro- and nanoscale, classical hydrodynamic
boundary conditions such as the no-slip condition no longer apply.
Instead, the flow profiles exhibit ``slip`` at the surface, which is
characterized by a finite slip length (partial slip). We present a
new, systematic way of implementing partial-slip boundary conditions
with arbitrary slip length in coarse-grained computer simulations.
The main idea is to represent the complex microscopic interface
structure by a spatially varying effective viscous force. An
analytical equation for the resulting slip length can be derived for
planar and for curved surfaces. The comparison with computer
simulations of a DPD (dissipative particle dynamics) fluid shows
that this expression is valid from full-slip to no-slip.
      }
%
%\begin{document}

\maketitle

\section{Introduction}
\label{sec:introduction}

In the last years much effort has been spent on the development of new microfluidic
devices. Recent success in reaching the nanometer scale has stimulated rising interest
in fluid mechanics on submicrometer scales. Coarse-grained computer simulations are
powerful theoretical tools for investigating the dynamical properties of such confined flows.
Simulation studies rely heavily on the adequate choice and implementation
of the boundary conditions at the surfaces. Traditionally, solid surfaces are
taken to be ``no-slip'' boundaries, {\em i.e.}, the fluid particles close to the
surfaces are assumed to be at rest relative to the surface. While this can be
rationalized to some extent for macroscopic (rough) surfaces~\cite{Einzel},
it has no microscopic justification. In fact, already J.~C.~Maxwell argued on
simple kinetic grounds that there always has to be some slip at surfaces, {\em i.e.},
the tangential velocity does not entirely vanish at the boundaries\cite{Maxwell,Kennard}.
In recent years, experiments have indicated that
the no-slip boundary condition is indeed usually not valid on the micrometer
scale\cite{Pit,Neto}. Instead, it has to be replaced by the ``partial-slip''
boundary condition
\begin{equation}
\label{eq:partial_slip}
\delta_B \: \: \partial_{{\bf n}} v_{\parallel}|_{{\bf r}_B} =  v_{\parallel}|_{{\bf r}_B},
\end{equation}
where $v_{\parallel}$ denotes the tangential component of the velocity and
$\partial_{{\bf n}} v_{\parallel}$ its spatial derivative normal to the surface, both
evaluated at the position ${\bf r}_B$ of the so-called ``hydrodynamic boundary''.
This boundary condition is characterized by two effective parameters, namely
(i) the slip length $\delta_B$ and (ii) the hydrodynamic
boundary ${\bf r}_B$. We note that the latter does not necessarily coincide
with the physical boundary. A boundary condition of the form
(\ref{eq:partial_slip}) can be formulated very generally, even in situations where
the ``physical boundary'' is not well-defined, {\em e.g.}, on surfaces covered
with a coating or a wetting layer. Eq.~(\ref{eq:partial_slip}) has been shown to
be the most general hydrodynamic boundary condition compatible with conservation laws
and irreversible thermodynamics\cite{Einzel}.

The question arises how to implement such partial-slip boundary
conditions in computer simulations. Despite the experimental
observation, most coarse-grained simulation models have aspired to
realize no-slip boundary conditions. This task already turned out to
be much more challenging than one might expect. For the dissipative
particle dynamics (DPD) simulation
method~\cite{Hoogerbrugge,Espanol}, several papers have been devoted
specifically to the problem of reaching no-slip at the boundaries.
Solid boundaries have been modeled, {\em e.g.}, by frozen fluid
regions\cite{Hoogerbrugge,Boek}, or by planar walls containing one
or several layers of embedded
particles\cite{Revenga,Duong-Hong,Pivkin}. It is worth noting that
the presence of frozen embedded particles alone does not warrant
no-slip -- they have to be supplemented by a bounce-back reflection
law at the walls. However, these may generate artifacts in the
temperature profiles\cite{Revenga}. Approaches to implementing arbitrary
partial-slip boundary conditions in Lattice-Boltzmann have recently
been proposed by Benzi \etal \cite{Benzi1,Benzi2}. However, to
our best knowledge, no such method has been formulated yet for 
coarse-grained particle simulations.

In this paper, we propose an alternative way of implementing
hydrodynamic boundary conditions in general coarse-grained
simulations (DPD and other). Compared to the above mentioned methods,
it has several advantages. (i) It does not use embedded or
frozen particles; hence, it does not introduce an artificial lateral
structure in the walls, and it is cheaper from a computational point
of view. (ii) It does not require the unphysical bounce-back
reflections and thus avoids their possible artifacts. (iii) The
basic idea is very general; it can be implemented for a wide range
of simulation methods in a very straightforward way. 
(iv) Most importantly: The method generates
full-slip, partial-slip as well as no-slip boundary conditions, and
it provides a model parameter $\alpha$ with which the slip length
can be tuned continuously and systematically from full-slip to
no-slip. Moreover, an analytic equation can be derived, which allows
to calculate the slip length to a very good approximation as a
function of $\alpha$. 

\section{The basic idea}
\label{sec:idea}

The hydrodynamic boundary conditions at the surface result from interactions
between the fluid particles and the walls. Depending on the microscopic structure of
the wall/fluid interface, these can be quite complex. In our approach, we replace
the boundaries by hard planar surfaces, and the unknown atomistic
forces by an effective coarse-grained friction force between the fluid particles
and the walls. This leads to a dissipation of the kinetic energy and therefore to
a decelerated velocity of the fluid close to the boundaries. The resulting
slip length depends on the strength of the friction force. At friction force zero,
one has a full-slip surface, {\em i.e.}, slip length infinity.
With increasing friction strength, the slip length decreases.

The idea can be applied very generally. In the following, we will focus on particle-based
off-lattice simulations. The friction force can then be implemented
by introducing spatially varying Langevin forces acting on particles $i$,
\begin{equation}
\label{eq:langevin}
  {\bf{F}}_i^{L}={\bf{F}}_i^{D}+{\bf{F}}_i^{R}
\end{equation}
The dissipative contribution
\begin{equation}
  {\bf{F}}_i^{D}=-\gamma_L\: \omega_L(z_i/z_c) \: \: ({\bf{v}}_i-{\bf{v}}_{wall})
\end{equation}
couples to the relative velocity $({\bf{v}}_i-{\bf{v}}_{wall})$ of the particle
with respect to the wall, with a locally varying viscosity $\gamma_L \omega_L(z_i/z_c)$
that depends on the wall-particle distance $z_i$, with a cutoff distance $z_c$.
The weighting function $\omega_L(\tau)$ is positive for $\tau < 1$ and zero
for $\tau > 1$. Otherwise, it can be chosen freely.
The prefactor $\gamma_L$ can be used to tune the strength of the friction
force and hence the value of the slip length.
To preserve the global temperature $T$ and to ensure the correct equilibrium
distribution, a random force obeying the fluctuation-dissipation relation
has to be added,
\begin{equation}
  F_{i,\alpha}^R= \sqrt{2 \gamma_L \: k_B T \: \omega_L(z_i/z_c)}\;\chi_{i,\alpha}
\end{equation}
with $\alpha = x,y,z$,
where $k_B$ is the Boltzmann constant and $\chi_{i,\alpha}$ a Gaussian distributed
random variable with mean zero and unit variance: $\langle \chi_{i,\alpha} \rangle = 0$,
$\langle \chi_{i,\alpha} \chi_{j,\beta} \rangle = \delta_{ij} \delta_{\alpha \beta}$.
Eq.~(\ref{eq:langevin}) can be used to model interactions with immobile
walls (such as channel boundaries) as well as interactions with surfaces of
mobile and/or rotating objects (such as colloids). In the latter case,
the force (\ref{eq:langevin}) must be balanced by a counterforce
$-{\bf{F}}_i^{L}$ and a countertorque $-(\bf{r}_i - \bf{R})\times {\bf{F}}_i^{L}$
acting on the object (${\bf R}$ being its center of mass).

An analytical expression for the slip length shall be derived in the
next section. In short, we find that the result can be written as a
function of the dimensionless parameter
\begin{equation}
\label{eq:alpha}
\alpha = z_c^2 \: \gamma_L \rho /\eta,
\end{equation}
where $\rho$ is the fluid density and $\eta$ its shear viscosity. To
leading order in $\alpha$, the slip length is given by
\begin{equation}
\label{eq:slip_planar}
 \frac{\delta_B}{z_c} = \frac{1}{\alpha
\int_0^1 {\rm d}\tau \: \omega(\tau)} + {\cal O}(\alpha^0).
\end{equation}
The sign of the next-to-leading correction is usually negative. For
example, a steplike weight profile $\omega(\tau) = 1$ for $\tau < 1$,
$\omega(\tau)=0$ otherwise, gives
\begin{equation}
\label{eq:db_step_expanded}
 \frac{\delta_B}{z_c} = \frac{1}{\alpha}
- \frac{2}{3} - \frac{1}{45} \alpha + \cdots.
\end{equation}
(cf. eq. (\ref{eq:db_step}) and a linear function $\omega(\tau)
=1-\tau$
\begin{equation}
\label{eq:db_linear_expanded}
 \frac{\delta_B}{z_c} = \frac{2}{\alpha}
- \frac{7}{15} - \frac{19}{1800} \alpha + \cdots
\end{equation}
(cf. eq. (\ref{eq:db_linear})). Hence the slip length becomes zero
for an appropriate choice of $\alpha$, and it can be tuned to any
value up to infinity (corresponding to $\alpha = 0$).

\section{Theory}
\label{sec:theory}

The theory is based on one crucial assumption: The Navier-Stokes
equations are taken to be valid on the length scale $z_c$ of the
cutoff. This assumption may seem bold, given that $z_c$ will
typically be chosen of the order of one particle diameter. However,
computer simulations~\cite{Alder,Duenweg} have shown that the
Navier-Stokes equations are valid down to surprisingly small length
scales. This is also supported by our own simulations (see below).

For simplicity, we begin with discussing planar surfaces. The
coordinate system is chosen such that the surface is at rest and
located at $z=0$. The stationary Stokes equation for our system
reads
\begin{equation}
\label{eq:stokes_planar}
\eta v''(z) = \rho \: \gamma_L \: \omega_L(z/z_c)  \: v(z) - \rho \: f_{ext}
\end{equation}
with the fluid density $\rho$ and the shear viscosity $\eta$. Here,
we assume that the fluid viscosity does not change in the vicinity
of the walls. The external force $f_{ext}$ incorporates, {\em e.g.},
the effect of a pressure gradient in the direction of flow. The
physical wall in our system is smooth, and we have no explicit
solid/fluid wall friction. Thus the shear stress in the fluid must
vanish, and the velocity field $v(z)$ satisfies the boundary
condition
\begin{equation}
\label{eq:boundary_0}
v'(z=0) = 0.
\end{equation}
We recall that the viscous force $\gamma_L \: \omega_L(z/z_c)  \: v(z)$ is only
active within a layer of finite thickness $z_c$. The total friction force
per surface area generated in this layer is given by
\begin{equation}
\label{eq:force}
F/A =  \rho \: \gamma_L \int_0^{z_c}  {\rm d}z \: \omega_L(z/z_c) \: v(z).
\end{equation}
Our goal is to replace the layer by an effective boundary between a
solid and an unperturbed fluid, {\em i.e.}, a hypothetical fluid not subjected
to the additional viscous forces in the layer.  The friction force per surface
area, $F/A$, is then equal to the frictional stress
on the solid, and to the shear stress on the unperturbed fluid at the position
$z=z_B$ of the boundary. This yields the effective boundary conditions
\begin{equation}
\label{eq:boundary}
-F/A
= \zeta_B v^{(0)}(z_B)
= \eta v^{(0)}{}'(z_B),
\end{equation}
where $v^{(0)}$ is the unperturbed velocity profile, and $\zeta_B$ the fluid-solid
friction coefficient. The comparison with eq. (\ref{eq:partial_slip}) shows that
the slip length can be identified with $\delta_B = \eta/\zeta_B$.

We first show that the hydrodynamic boundary must be identical with the
physical boundary,
\begin{equation}
\label{eq:zb}
z_B = 0.
\end{equation}
To this end, we first insert eq. (\ref{eq:stokes_planar}) in eq. (\ref{eq:force}) and
perform a partial integration, taking advantage of the boundary condition
(\ref{eq:boundary_0}), to obtain
\begin{equation}
-F/A = \eta v'(z_c) + \rho f_{ext} z_c.
\label{eq:f1}
\end{equation}
According to eq.~(\ref{eq:boundary}), this must be equal to $\eta v^{(0)}{}'(z_B)$.
The unperturbed velocity profile $v^{(0)}$ solves the Stokes equation
at $\gamma_L = 0$ with the boundary conditions
\begin{equation}
\label{eq:boundary_c}
v^{(0)}(z_c) = v(z_c); \quad
v^{(0)}{}'(z_c) = v'(z_c).
\end{equation}
In the planar case (\ref{eq:stokes_planar}), one obtains
\begin{equation}
\label{eq:v0_planar}
v^{(0)}(z) = v(z_c) + v'(z_c) (z-z_c)
- \frac{\rho}{2 \eta} f_{ext} (z - z_c)^2
\end{equation}
and hence
\begin{equation}
\label{eq:dv0_planar} \eta v^{(0)}{}'(z_B) = \eta v'(z_c) + f_{ext}
\rho (z_c - z_B).
\end{equation}
Comparing eqs.~(\ref{eq:dv0_planar}) and (\ref{eq:f1}) gives eq.~(\ref{eq:zb}).

Next we calculate the slip length. This is most conveniently done
for the case $f_{ext}=0$. We first integrate
eq.~(\ref{eq:stokes_planar}) for given $v(0) =: v_0$ and $v'(0) = 0$
(eq.~(\ref{eq:boundary_0})) to get $v(z_c)$ and $v'(z_c)$. {\em Via}
eq.~(\ref{eq:v0_planar}), we then determine the values
$v^{(0)}(z_B)$ and $v^{(0)}{}'(z_B)$ of the unperturbed profile at
the hydrodynamic boundary. This finally allows to calculate the slip
length from 
\begin{equation}
\label{eq:sliplength_final}
\delta_B = v^{(0)}(z_B)/v^{(0)}{}'(z_B) 
\end{equation}
(cf.  eq.~(\ref{eq:boundary})). We note that all profiles scale linearly
with $v_0$, hence the final expression for the slip length does not
depend on $v_0$ any more. More generally, the result depends only on
the dimensionless quantity $\alpha$ defined in eq. (\ref{eq:alpha}),
in the sense that the slip length in units of $z_c$, {\em i.e.}, the
quantity $\delta_B/z_c$, can be written as a function of $\alpha$
only. For example, if the weight function $\omega_L(\tau)$ is a
simple step function, $\omega_L(\tau) = 1$ for $\tau < 1$, the flow
profile $v(z)$ at $z < z_c$ is a superposition of exponentials
$\exp(\pm \sqrt{\alpha} z)$, and one obtains
\begin{equation}
\label{eq:db_step} \frac{\delta_B}{z_c} = \frac{1}{\sqrt{\alpha}
\tanh(\alpha)} - 1.
\end{equation}
If $\omega_L(\tau)$ drops linearly to zero, $\omega_L(\tau) = 1 -
\tau$, the profile can be written as (see appendix A)
\begin{equation}
\label{eq:db_linear} \frac{\delta_B}{z_c} =
  -1+\;\frac{1}{(3\alpha)^{1/3}}\;
  \frac{\Gamma\left(\frac{1}{3}\right)}{\Gamma\left(\frac{2}{3}\right)} \;
  \frac{I_{-2/3}\left(\frac{2\sqrt{\alpha}}{3}\right)}
  {I_{2/3}\left(\frac{2\sqrt{\alpha}}{3}\right)},
\end{equation}
where $\Gamma$ is the Gamma-Function and $I$ the modified Bessel
function of the first kind.

The slip length can become zero (corresponding to no-slip) or even
negative. The no-slip boundary condition is obtained at $\alpha =
1.433$ for a steplike weight function, and at $\alpha = 3.973$ for a
linear weight function. Negative slip lengths are encountered at
even larger $\alpha$. In this case, the hypothetical unperturbed
profile $v^{(0)}$ changes sign close to the boundary. We note that
the true velocity profile, $v(z)$, never changes sign; hence our
negative slip lengths do not correspond to unphysical situations.

In the regime of positive slip lengths, $\alpha$ is small and can be
used as an expansion parameter. Expanding eqs. (\ref{eq:db_step})
and (\ref{eq:db_linear}) in powers of $\alpha$ gives
(\ref{eq:db_step_expanded}) and (\ref{eq:db_linear_expanded}). More
generally, one can derive a useful expression for arbitrary weight
functions. We first note that the true velocity profile and the
unperturbed profile are identical at the order $\alpha^0$, {\em
i.e.}, $v(z) = v^{(0)}(z) + {\cal O}(\alpha)$. Furthermore, the
derivative of the velocity profiles is of order $\alpha$, by virtue
of eqs.~(\ref{eq:force}), (\ref{eq:boundary}), and
(\ref{eq:dv0_planar}), hence one even has $v(z) = v^{(0)}(z_B) +
{\cal O}(\alpha)$. Applying once more eqs.~(\ref{eq:force}),
(\ref{eq:boundary}), and (\ref{eq:v0_planar}), we obtain eq.
(\ref{eq:slip_planar}). This equation allows to estimate the slip
length reasonably accurately for arbitrary choices of the weight
function $\omega(\tau)$.

\begin{figure}[t!]
  \includegraphics{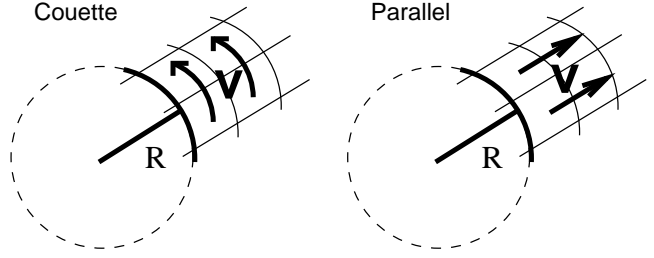}
  \caption{Two types of cylindrically curved geometries. In the Couette geometry (left),
  the flow profile bends along the curved surface. In the parallel geometry, the fluid
  flows in the perpendicular direction.
    }
  \label{fig:curved}
\end{figure}
The extension of the theory to curved geometries is straightforward. Here we discuss
cylindrical curvature ({\em i.e.}, edges). One has to distinguish between
curvature in the direction of flow (the ``Couette'' case) and curvature in the transverse
direction (the ``parallel'' case), see Fig. \ref{fig:curved}.  The curvature is characterized
by the radius $R$ of a tangent sphere, which we take by convention to be
positive at the surface of convex objects, and negative at the surface of
concave objects. Details of the calculation are given in the appendix B.
As in the planar case, the hydrodynamic boundary is found to be identical
with the physical boundary. To leading order of $\alpha$, the slip length
is given by
\begin{equation}
\label{eq:slip_couette}
\frac{z_c}{\delta_B}
= \frac{z_c}{R} +
  \alpha \int_0^1 {\rm d}\tau \: \omega(\tau) \: (1 + \tau \frac{z_c}{R})^2
  + {\cal O}(\alpha^0).
\end{equation}
in the Couette case, and
\begin{equation}
\label{eq:slip_parallel}
\frac{z_c}{\delta_B}
= \alpha \int_0^1 {\rm d}\tau \: \omega(\tau) \: (1 + \tau \frac{z_c}{R})
  + {\cal O}(\alpha^0).
\end{equation}
in the parallel case. In the Couette case, the inverse slip length has a contribution
$\frac{1}{R}$ of purely geometric origin. The existence of this term
has been pointed out by Einzel \etal\cite{Einzel}. The remaining term can be
identified with a ``microscopic'' inverse slip length.  Both in the Couette and the
parallel case, this microscopic slip length is reduced at the surface of convex objects,
and enhanced at the surface of concave objects.

\section{Comparison with computer simulations}
\label{sec:simulations}

To test the method and the theory, we have carried out dissipative
particle dynamics\cite{Hoogerbrugge,Espanol} (DPD) simulations of
fluids confined in a parallel slit.  The fluid particles interact
with purely dissipative DPD forces of range $\sigma$~\cite{Espanol}.
They have no conservative interactions with each other, their static
equilibrium structure is thus that of an ideal gas. They are
confined into the slit by two parallel walls, with which they
interact {\em via} a Weeks-Chandler-Andersen potential~\cite{WCA}
with characteristic length $\sigma$; alternatively, one can use hard
reflecting walls. In the vicinity of the walls, up to the cutoff
distance $z_c$, the particles experience a wall friction force of
the form (\ref{eq:langevin}) with a linear weight function,
$\omega_L(\tau) = 1 - \tau$. The choice of the weight function is in
fact completely arbitrary. It was motivated by the fact that the
DPD weight factors for intermolecular dissipative interactions are
also usually chosen linear for reasons of computational efficiency.
The natural units of the simulation are the length unit $\sigma$, 
the thermal energy $k_B T =:\epsilon$, and
the mass of the particles $m$. In these units, the cutoff at the
walls was chosen $z_c = \sigma$, and the time step $\delta t = 0.01
\: \sigma \sqrt{m/\epsilon}$. Periodic boundary conditions were
applied in the slit plane. All simulations were carried out with
extensions of the simulation package {\sf
ESPResSo}\cite{Espresso1,Espresso2,Espresso3}.

\begin{figure}[t!]
  \includegraphics{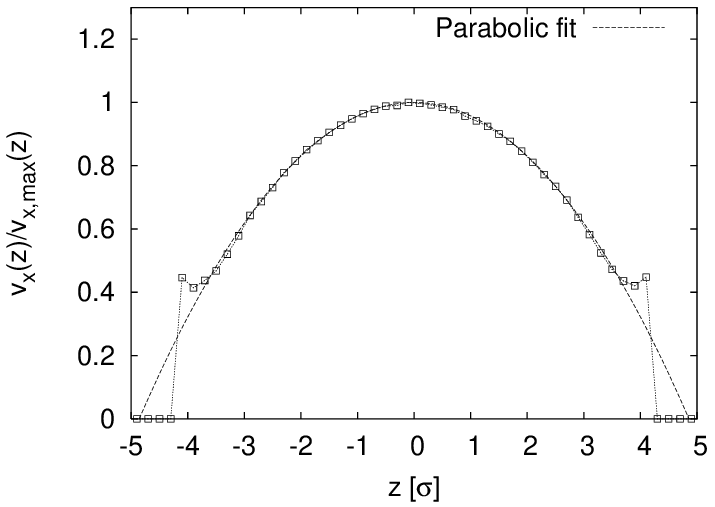}
  \includegraphics{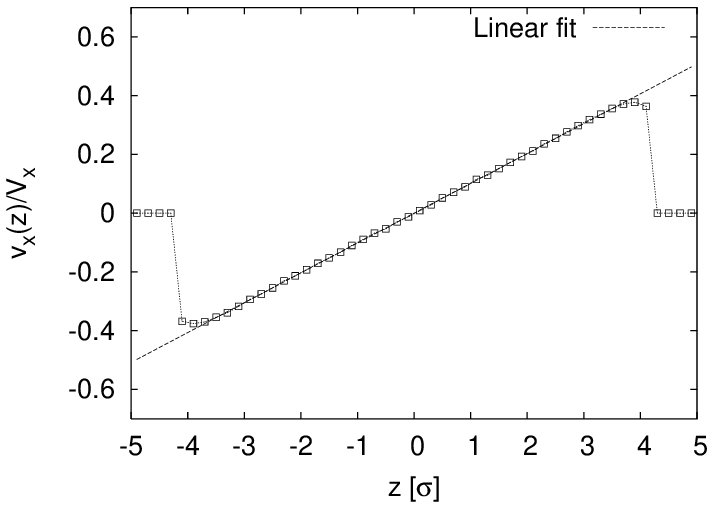}
  \caption{Normalised plane Poiseuille flow (top) and plane Couette flow (bottom)
   for DPD fluids ($\gamma_{DPD} = 5 \sqrt{m \epsilon}/\sigma$) with number density
   $\rho = 3.75 \sigma^{-3}$. The parameters of the surface friction force are
   $\gamma_L=1.0 \sqrt{m \epsilon}/\sigma$, $z_c = \sigma$.
   The dashed lines show fits to the theoretical profiles.
    }
  \label{fig:Poiseuille_Couette}
\end{figure}
We have simulated both planar Couette and planar Poiseuille flow\cite{Lu}. In the first
case, one of the walls moves at constant speed relative to the other. In the second
case, the effect of the pressure gradient is mimicked by a constant force
acting on all particles in the slit.  Fig.~\ref{fig:Poiseuille_Couette} shows two examples
of flow profiles. Sufficiently far from the walls, the classical parabolic Poiseuille
flow and linear Couette flow profiles are nicely recovered.

The combined results from the Poiseuille and Couette fits allows to calculate
the slip length and the position of the hydrodynamic boundary independently:
We determine the distance $P$ between the two points where the
extrapolated Poiseuille parabola vanishes, and the distance $C$ between the two
points where the extrapolated Couette line reaches the velocities of the two walls.
The slip length is then given by
\begin{equation}
\delta_B^2 = (C^2 - P^2)/4,
\end{equation}
and the hydrodynamic boundary is located at the distance
\begin{equation}
z_B = \delta_B - (C-L)/2.
\end{equation}
from the physical boundary, where $L$ is the distance between the two physical walls.
We note that both $C$ and $P$ enter the expression for the slip length,
hence both the Poiseuille and the Couette profile have to be simulated.
This is because we do not make any assumptions regarding the hydrodynamic 
boundary. If its position is unknown, two types of profiles have to
be measured in order to determine the two parameters $z_B$ and
$\delta_B$. This fact is not always appreciated in the literature.

\begin{figure}[t!]
  \includegraphics{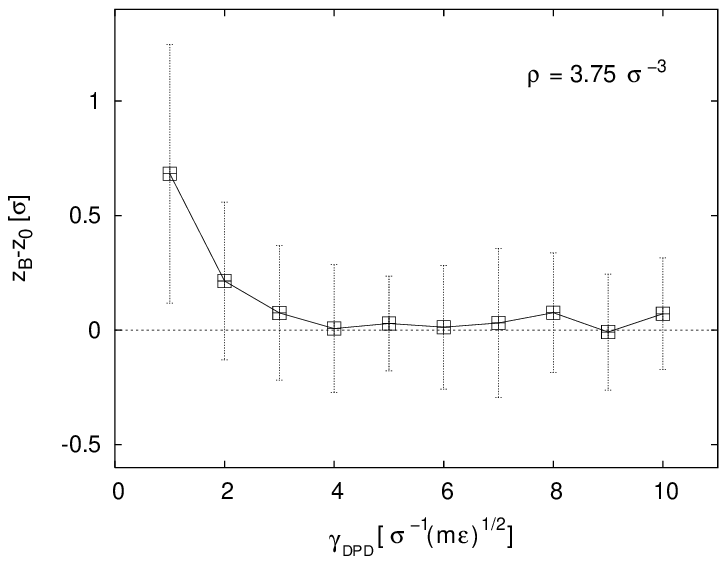}
  \includegraphics{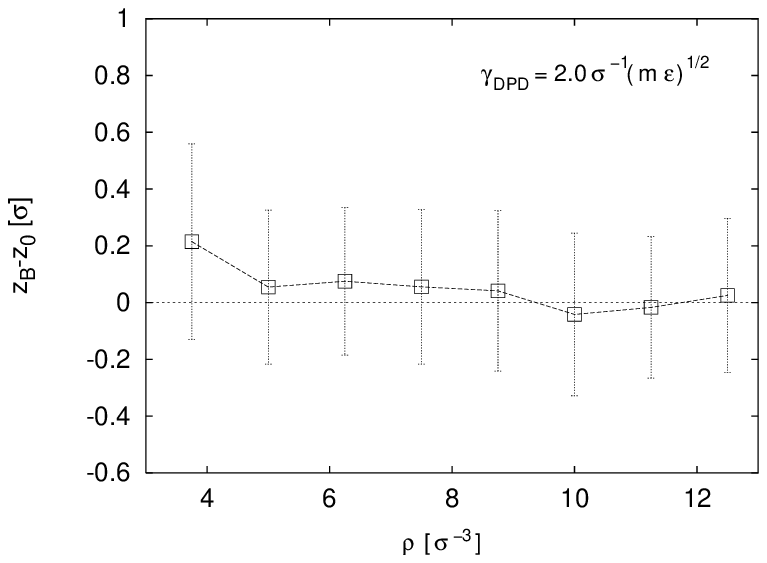}
  \caption{Position of the hydrodynamic boundaries for varying $\gamma_{DPD}$
    and fixed number density $\rho=3.75 \sigma^{-3}$ (top)
    and for varying rho and fixed $\gamma_{DPD} = 2 \sqrt{m \epsilon}/\sigma$ (bottom).
    The result does not depend on the value of the surface friction $\gamma_L$.
    The data shown here are averaged over all $\gamma_L$.
    }
  \label{fig:boundary}
\end{figure}
\begin{figure}[t!]
  \includegraphics{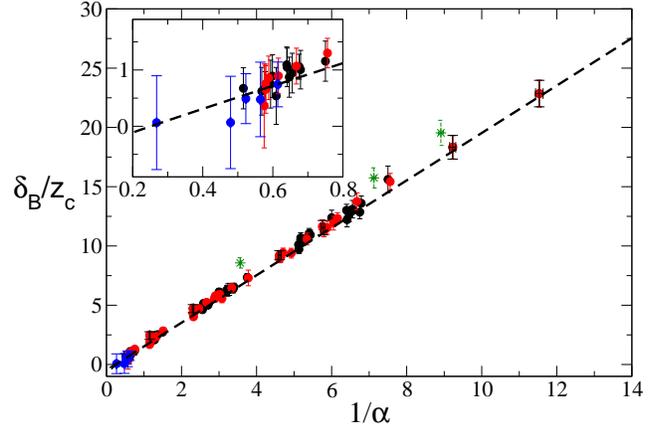}
  \caption{Slip length $\delta_B$ in units of $z_c$ vs.\ $\alpha$ for varying values of the
   parameter triplet $(\rho,\gamma_{DPD},\gamma_L)$ (in units of $\sigma^{-3}$ or
   $\sqrt{m \epsilon}/\sigma$, respectively).
    Black: series with $\rho$ fixed: (3.75 , 2-10, 0.1-1).
    Red: series with $\gamma_{DPD}$ fixed: (3.75-12.5, 2, 0.1-1).
    Blue: selected triplet values: (6.35,5,1),(5,5,1), (11.25,2,1.1), (11.25,2,1.2), (3.75,10,2.5).
    Green star symbols: series corresponding to the point in fig.~\protect\ref{fig:boundary},
    where the position of the hydrodynamic boundary deviates from theoretical expectation: (3.75,1,0.1-1).
    Dashed black line: Theoretical prediction of eq.~(\protect\ref{eq:db_linear}).
    The inset shows a blowup of the same data.
     }
    \label{fig:slip_theory}
\end{figure}

We have determined the hydrodynamic boundary and the slip length for the range
of parameters $\gamma_L = (0.1$ -$ 5) \:\sqrt{m \epsilon}/\sigma$ (the friction at the wall),
$\rho = (3.75$-$12.5) \:\sigma^{-3}$ (the number density of fluid particles), and
$\gamma_{DPD} = (1$-$10) \sqrt{m \epsilon}/\sigma$
(the friction coefficient of the DPD forces).  The results are summarized in the
figures \ref{fig:boundary} and \ref{fig:slip_theory}. The theory predicts that
the hydrodynamic boundary is identical with the physical boundary, eq.~(\ref{eq:zb}).
This was found to be correct for most systems. Significant deviations were only
encountered at small DPD-friction $\gamma_{DPD}$ (fig.~\ref{fig:boundary}).
Fig.~\ref{fig:slip_theory} shows the results for the slip length as a function
of the dimensionless parameter $\alpha = \gamma_L \rho z_c^2/\eta$. The shear
viscosity $\eta$ of the fluids was determined from the Poiseuille fits\cite{Backer}.
The numerical results for the slip length compare very favorably with the theoretical
prediction of eq.~(\ref{eq:db_linear_expanded}), except in the regime where the
hydrodynamic boundary also deviates from its theoretical position, {\em i.e.}, at
small DPD-friction $\gamma_{DPD}$. Even in this regime, the method can still be
used to produce partial slip. A wide range of slip lengths has been realized.
The inset of Fig.~\ref{fig:slip_theory} illustrates that the no-slip regime
can also be reached very naturally.

\section{Summary}
\label{sec:summary}

We have presented a method that allows to implement arbitrary
hydrodynamic boundary conditions in coarse-grained simulations. We have
illustrated and tested this method with DPD simulations; however
it can be applied much more generally. The main idea is to introduce
an effective surface friction force, complemented by an appropriate thermostat.
This idea can also be implemented, {\em e.g.}, in stochastic rotation dynamics
simulations or in Lattice Boltzmann simulations. The strength of the 
friction force effectively determines the slip length. We have shown that 
our method allows one to tune the slip length systematically from
full-slip to no-slip. Furthermore, we have derived an analytical equation
for the slip length as a function of the strength of the friction force and
the shear viscosity of the fluid.

An alternative force-driven method to control the slip length on
surfaces has recently been proposed for the Lattice Boltzmann
simulation method by Benzi et al\cite{Benzi1}. In this work,
conservative forces are introduced that deplete the fluid density at 
the surface, which leads to partial slip -- a physical effect. 
In contrast, the dissipative forces in our method mimick directly the 
effective friction between the surface and the bulk fluid. 
While the method does thus not contribute 
to a microscopic understanding of the slip at surfaces, it does offer 
a way to model the essential
wall properties, from the hydrodynamics point of view, and to absorb
the complex wall structure into a well-defined coarse-grained
parameter set\cite{Benzi2}. The method can be implemented easily, it
generates the partial-slip boundary conditions for microfluidic
simulations that are observed in experiments. Curvature can be
introduced in a straightforward manner, the effect of curvature on
the slip length in the model can be calculated analytically, and
corrected by an easy change of local friction if necessary. Hence
the method is suitable to study complex geometries and/or objects,
such as structured channels, channels with objects, or rotating
nanoparticles in flow.

The only obvious restriction is that the characteristic length scale
of the structures has to be larger than the length scale $z_c$ on
which the friction force is applied. In future work, it will be
interesting to determine the minimal lateral length scale that the
method can handle, {\em e.g.}, by studying surfaces with mixed
boundary conditions. For example, analytical results are available
for the macroscopic effective slip length on surfaces covered by
stripes with alternating full-slip and no-slip boundary
conditions\cite{Philip,Lauga}, and the Stokes equation has also been
solved numerically for more general mixed
surfaces\cite{Cottin-Bizonne,Priezjev}. This will be a good starting
point for a systematic study. Furthermore, we plan to apply our
method to the simulation of electrolyte flow to assess the influence
of partial-slip boundary conditions on electroosmotic
flow\cite{Smiatek}.

%\acknowledgments

We thank Burkhard D\"{u}nweg, Christian Holm and Ulf D.~Schiller for
nice and fruitful discussions.
This work was funded by the Volkswagen Stiftung.
The visit of MPA to Bielefeld was supported by the Alexander von Humboldt foundation.

\section{Appendices}
\label{sec:appendix}

\setcounter{equation}{1}
\renewcommand{\theequation}{A.\arabic{equation}}

\subsection*{A: Slip length for linear weight function}
\label{sec:appendixa}

The program for calculating the slip length (Eq.~(\ref{eq:sliplength_final}) 
can easily be carried out numerically for
arbitrary weight functions $\omega(\tau)$. For some simple choices of
the weight function, analytical results can also be obtained.
To illustrate the approach, we sketch the calculation leading to 
Eq.~(\ref{eq:db_linear}),
the expression for the slip length for a linear weight function
$\omega(\tau) = 1-\tau$.

In the absence of an external force ($f_{ext}=0$), the general solution of
Eq.~(\ref{eq:stokes_planar}) for a linear weight function reads
\begin{equation}
\label{eq:vzapp}
v(z) = A \: \mbox{Ai}(c (1-z/z_c)) + B \: \mbox{Bi}(c (1-z/z_c))
\end{equation}
with $c=-(- \alpha)^{1/3}$, where Ai and Bi are the Airy
functions and $\alpha$ has been defined in Eq.~(\ref{eq:alpha}).
The boundary conditions $v(0)=v_0, v'(0)=0$ determine the coefficients
\begin{eqnarray*}
\label{eq:vzappA}
A &=& v_0 \mbox{Ai}'(c)
/ (\mbox{Ai}'(c) \mbox{Bi}(c) - \mbox{Ai}(c) \mbox{Bi}'(c)) \\
B &=& v_0 \mbox{Bi}'(c)
\label{eq:vzappB}
/ (\mbox{Ai}'(c) \mbox{Bi}(c) - \mbox{Ai}(c) \mbox{Bi}'(c)).
\end{eqnarray*}
The unperturbed profile is linear,
$v^{(0)}(z) = v(z_c) + v'(z_c) (z-z_c)$, and the hydrodynamic boundary
is located at $z_B=0$ according to Eq.~(\ref{eq:zb}), hence the
equation (\ref{eq:sliplength_final}) for the slip length can be written as
\begin{equation}
\frac{\delta_B}{z_c} = \frac{v(z_c)}{z_c v'(z_c)}-1.
\end{equation}
Inserting this in Eqs.~(\ref{eq:vzapp}) with
(\ref{eq:vzappA}) and (\ref{eq:vzappB}) and using the identities
$\mbox{Ai}(0) = 1/(3^{2/3} \Gamma(\frac{2}{3}))$, 
$\mbox{Ai}'(0) = - 1/(3^{1/3} \Gamma(\frac{1}{3}))$,
$\mbox{Bi}(0) = \sqrt{3} \mbox{Ai}(0)$, and
$\mbox{Bi}'(0) = - \sqrt{3} \mbox{Ai}'(0)$, 
where $\Gamma$ is the Euler gamma function,
one obtains
\begin{equation}
\label{eq:db_linear_app}
\frac{\delta_B}{z_c} = -1 - \frac{1}{3^{1/3} c} \:
 \frac{\Gamma(\frac{1}{3})}{\Gamma(\frac{2}{3})} \:
\frac{\mbox{Ai}'(c) \sqrt{3} - \mbox{Bi}'(c)}
{\mbox{Ai}'(c) \sqrt{3} + \mbox{Bi}'(c)}.
\end{equation}
To get rid of the complex argument $c$, we use the series expansion of
the Airy functions\cite{wolfram}. The functions of interest for us can 
be written as $\mbox{Ai}'(z) \sqrt{3} = I_1 - I_2$ and
$\mbox{Bi}'(z) = I_1 + I_2$ with
\begin{eqnarray*}
I_1 & = &\frac{z^2}{3 \: 3^{1/6}}
\sum_k \frac{1}{\Gamma(k + 5/3)\: k!} \: (\frac{z^3}{9})^k \\
I_2 & = & 3^{1/6}
\sum_k \frac{1}{\Gamma(k + 1/3)\: k!} \: (\frac{z^3}{9})^k.
\end{eqnarray*}
Comparing this with the series representation of the modified Bessel
function of the first kind,
\begin{equation}
I_{\nu}(z) = (\frac{z}{2})^\nu
\sum_k \frac{1}{\Gamma(k+\nu+1) \: k!} \: (\frac{z}{2})^{2 k},
\end{equation}
one easily verifies the identities
\begin{displaymath}
I_1  =  - \frac{\alpha^{1/3}}{\sqrt{3}} \: I_{-2/3}(\frac{2
\sqrt{\alpha}}{3})  \quad \mbox{and} \quad
I_2 =  \frac{\alpha^{2/3}}{\sqrt{3} c} \: I_{2/3}(\frac{2
\sqrt{\alpha}}{3}).
\end{displaymath}
Inserting this into Eq.~(\ref{eq:db_linear_app}) gives
Eq.~(\ref{eq:db_linear}).

\subsection*{B: Curved boundaries}
\label{sec:appendixb}

The strategy for calculating the slip length and the hydrodynamic boundary position at
curved surfaces is perfectly analogous to that sketched in the main
text for the planar geometry.
We begin with noting that the expression for the total friction force per surface area,
eq.~(\ref{eq:force}), acquires an additional geometric factor at curved surfaces,
\begin{equation}
\label{eq:force_curved}
F/A =  \rho \gamma_L \int_R^{R+z_c} {\rm d}r \: \omega_L((r-R)/z_c) \: \: v(r)
\: \frac{r}{r_B},
\end{equation}
which accounts for the fact that the viscous layer is stretched or compressed in the
presence of curvature. Here, $R$ is the radius of curvature, $r$ the distance from the
center of the tangent sphere, and $r_B$ the position of the hydrodynamic boundary.
Furthermore, the Stokes equation, (\ref{eq:stokes_planar}), and the boundary conditions,
(\ref{eq:boundary_0}) and (\ref{eq:boundary}) have to be replaced by the appropriate
cylindrical versions.

We first discuss the Couette case. The Stokes equation for this geometry reads
\begin{equation}
\label{eq:stokes_couette}
\eta \: \partial_r \frac{1}{r} \partial_r r v(r) =
\rho \: \gamma_L \omega( (r-R)/z_c) \: v(r) - \rho \: f_{ext},
\end{equation}
The unperturbed solution of
this equation for $\gamma_L = 0$ and fixed $v(r_c) = v_c, v'(r_c)=v_c'$
is given by
\begin{equation}
\label{eq:v0_couette}
v^{(0)}(r) = \frac{1}{2} \big[
\frac{r}{r_c} (v_c + v_c' r_c + \tilde{f})
+ \frac{r_c}{r} (v_c - v_c' r_c - \frac{\tilde{f}}{3})
- \frac{\tilde{f}}{3} (\frac{r}{r_c})^2 \big]
\end{equation}
with $\tilde{f} = \rho f_{ext} r_c^2/\eta$. In the absence of an external force,
$\tilde{f}=0$, one recovers the well-known profile of cylindrical Couette flow.
The appropriate expressions for the fluid shear stress
also differ from those in the planar case, and Eqs. (\ref{eq:boundary_0})
and (\ref{eq:boundary}) have to be replaced by\cite{Einzel}
\begin{equation}
\label{eq:boundary_couette_0}
( \partial_r v - v/r ) |_{r = R} = 0,
\end{equation}
\begin{equation}
\label{eq:boundary_couette}
-F/A
= \zeta_B v^{(0)}(r)|_{r = r_B}
= \eta ( \partial_r v^{(0)} - v^{(0)}/r ) |_{r = r_B}.
\end{equation}
Comparing eq.~(\ref{eq:boundary_couette}) with eq.~(\ref{eq:partial_slip}), one immediately
obtains $\delta_B^{-1} = \delta_0^{-1} + r_B^{-1}$ with $\delta_0^{-1} = \zeta_B/\eta$,
where $r_B^{-1}$ is a geometric contribution, and $\delta_0^{-1}$ incorporates the
specific effect of the surface structure. Comparing with eq.~(\ref{eq:force_curved}),
one gets
\begin{equation}
\label{eq:d0}
\frac{z_c}{\delta_0}
 =  \alpha \int_0^1 {\rm d}\tau \: \omega_L(\tau)
\: \frac{v(R+\tau z_c)}{v^{(0)}(r_B)}
\: \frac{R+\tau z_c}{r_B}.
\end{equation}
For fixed $v(R+z_c)=v_c, v'(R+z_c)=v_c'$, the actual profile $v(r)$ is identical
with the unperturbed profile $v^{(0)}(r)$ of eq.~(\ref{eq:v0_couette}) at
leading order of $\alpha$. Inserting this in eq.~(\ref{eq:d0}) and using
$r_B = R$, we finally recover eq.~(\ref{eq:slip_couette}).
To establish the identity $r_B = R$, we require again that the total
friction force as given by eq.~(\ref{eq:force_curved}) is identical with the
hypothetical shear stress on the unperturbed fluid, eq.~(\ref{eq:boundary_couette}).
Inserting (\ref{eq:stokes_couette}) in (\ref{eq:force_curved}), carrying out
a few partial integrations, using (\ref{eq:boundary_couette_0}), and
finally replacing once more $v(r)$ by $v^{(0)}(r)$ at order $\alpha$, we obtain
\begin{equation}
- \frac{F}{A} =
\frac{\eta}{r_B} \frac{r_c}{R} \big\{ -v_c + v_c' r_c
+ \frac{\tilde{f}}{3} (1 + (\frac{R}{r_C})^3) \big\}
+ {\cal O}(\alpha).
\end{equation}
Comparing this with eq.~(\ref{eq:boundary_couette}), after inserting (\ref{eq:v0_couette}),
we find that $r_B$ must be equal to the surface position $R$ at the leading order of $\alpha$.

The parallel case is even simpler. The appropriate Stokes equation reads
\begin{equation}
\label{eq:stokes_parallel}
\eta \frac{1}{r} \partial_r r \partial_r v(r) =
\rho \: \gamma_L \omega( (r-R)/z_c) \: v(r) - \rho \: f_{ext}.
\end{equation}
In the unperturbed case $\gamma_L = 0$, this is solved by
\begin{equation}
\label{eq:v0_parallel}
v^{(0)}(r) = v_c + (v_c' r_c + \frac{\tilde{f}}{2}) \ln(\frac{r}{r_c}) +
\frac{\tilde{f}}{4} (1-(\frac{r}{r_c})^2)
\end{equation}
with $v_c := v(r_c)$, $v_c' := v'(r_c)$, and $\tilde{f} = \rho f_{ext} r_c^2/\eta$ as before.
The expressions for the fluid shear stress are similar to those in the planar case,
\begin{equation}
\label{eq:boundary_parallel_0}
 \partial_r v |_{r = R} = 0,
\end{equation}
\begin{equation}
\label{eq:boundary_parallel}
-F/A
= \zeta_B v^{(0)}(r)|_{r=r_B}
= \eta ( \partial_r v^{(0)} - v^{(0)}/r ) |_{r = r_B}.
\end{equation}
Proceeding as before, we find that $r_B = R$ is fulfilled exactly,
and we recover eq.~(\ref{eq:slip_parallel}).

\end{document}